# Neural Network Image Reconstruction for Magnetic Particle Imaging

Byung Gyu Chae

We investigate neural network image reconstruction for magnetic particle imaging. The network performance depends strongly on the convolution effects of the spectrum input data. The larger convolution effect appearing at a relatively smaller nanoparticle size obstructs the network training. The trained single-layer network reveals the weighting matrix consisted of a basis vector in the form of Chebyshev polynomials of the second kind. The weighting matrix corresponds to an inverse system matrix, where an incoherency of basis vectors due to a low convolution effects as well as a nonlinear activation function plays a crucial role in retrieving the matrix elements. Test images are well reconstructed through trained networks having an inverse kernel matrix. We also confirm that a multi-layer network with one hidden layer improves the performance. The architecture of a neural network overcoming the low incoherence of the inverse kernel through the classification property will become a better tool for image reconstruction.

This work was supported by Institute for Information & communications Technology Promotion (IITP) grant funded by the Korea government (MSIP) [B0132-15-1001, Development of Next Imaging System: XIS].

Byung Gyu Chae (bgchae@etri.re.kr) is with the Biomedical IT Research Laboratory, ETRI, Daejeon, Rep. of Korea.

## I. Introduction

Magnetic particle imaging (MPI) is a new imaging modality capable of observing the spatial distribution of superparamagnetic particles with a strong nonlinear magnetic response [1]-[5]. Image reconstruction is generally achieved by solving the linear equation with the system function, which maps the measured signal spectrum into the spatial concentration [3], [4]. The system function is not simply obtained from the geometry property, as with the kernels for computerized tomography and magnetic resonance imaging. The function includes nonlinear magnetic responses of the particles as well as the measurement conditions, and can be interpreted as a convolution of orthogonal special functions with the derivative of the Langevin function [6].

Image reconstruction in MPI uses the kernel matrix made through a direct measurement or a model-based simulation where an ill-posed inverse problem may be inevitable because of a large convolution effect appearing at a relatively smaller nanoparticle size, or because of external noise [7]. An analytic method does not apply to the image reconstruction directly, and therefore, a regularization technique is necessary to retrieve an image with high resolution. Various iterative methods [8]-[11] have been used for a successful reconstruction of a particle image. However, such iterative methods have a problem in terms of the large amount computational cost incurred when compared to an analytic technique.

A neural network technique [12]-[16] has recently been studied for image reconstruction in computerized tomography. The neural network maps the input data to the output data through a nonlinear activation function. An input of even a small amount of projections is well trained by a multi-layer network, and the trained network easily retrieves an image of the test samples. This method has a good merit in that, once the network is learned, the image is easily reconstructed through a forward propagation of the trained networks, which can reduce the computational load significantly. A neural network approach for MPI has been attempted by using input data in the temporal domain [17], but no detailed studies utilizing the spectrum data have yet been conducted.

In this work, we carry out image reconstruction using a neural network technique with MPI spectrum data. The spectrum data used as inputs are acquired from a model-based simulation for randomly distributed magnetic particles. These supervised data are trained through the adopted neural networks consisting of a simple feedforward structure. The learning behavior and network evaluation are investigated in detail under various network conditions. In particular, we explored the strong dependence of the network performance on the magnetic particle size. We then analyze the convolution effects of the spectral input data on the network learning, and herein describe the relation between the inverse system kernel and weighting matrix.

## II. Neural Network Image Reconstruction for MPI

Ferromagnetic nanoparticles with a single domain behave as if in a paramagnetic state for an external magnetic field because of a relatively low relaxation time, whereas a magnetic susceptibility much larger than that of a conventional paramagnet is maintained, which is called super-paramagnetism [3]. The MPI apparatus images the biological region of interest involving the nanoparticles used as a contrast agent such that the magnetic response of the particle is measured. The magnetic response signal of a particle located at a particular position is acquired by detecting the nonlinear magnetization response to a sinusoidal drive field in the region of the field-free point (FFP) of a magnetic field.

The spatial area is effectively measured by moving the FFP along the Lissajous trajectory, which is composed of voxels [4], [6]. The gradient field added to the drive field enables scanning the field of view, and thus a convolved signal with a particle distribution is obtained. Considering a one-dimensional field scan, the total magnetic field is given by

$$H(x,t) = Gx + A_D \cos(2\pi f t), \quad (1)$$

where $G$ is the gradient strength, $A_D$ is the drive field amplitude, and $f$ is the frequency of the drive field. The induced voltage in a receive coil with homogeneous sensitivity $p_s$ for the particle distribution $c(x)$ is expressed as

$$u(t) = -\mu_0 \, p_s \int_\Omega \frac{\partial M(x,t)}{\partial t} c(x) dx. \quad (2)$$

Here, $\mu_0$ is the magnetic permeability of a vacuum. The magnetization $M(H)$ is described as a Langevin function of the applied magnetic field. We found that equation (2) indicates the convolution of the derivative of the magnetization with the particle concentration. The signal spectrum in a frequency domain is represented as:

$$u_k = -\mu_0 \, p_s \int_\Omega \left( \int_0^T \frac{\partial M(x,t)}{\partial t} \exp\left(\frac{i2\pi k t}{T}\right) dt \right) c(x) dx. \quad (3)$$

The above equation is rewritten in a matrix form using a system matrix $\mathbf{S}$ as follows:

$$\mathbf{u} = \mathbf{S}\mathbf{c}. \quad (4)$$

Once the inverse system matrix is calculated, the particle concentration $\mathbf{c}$ with respect to the measured signal spectrum vector $\mathbf{u}$ can be obtained. However, the system matrix is relatively complex unlike a simple geometric kernel, which implies both dynamical measurement conditions and magnetic system parameters. This inverse problem is solved using a regularization technique to relax the ill-conditioning of the matrix [8], [10] as follows:

$$\arg\min_{\mathbf{c}} \|\mathbf{S}\mathbf{c} - \mathbf{u}\|_2^2 + \lambda \|\mathbf{c}\|_p, \quad (5)$$

where $\lambda$ is a regularization parameter and $p$ indicates the Lp-norm. Several iterative methods find the optimal value of the particle information updated during a repetitive process.

Single-layer neural networks make it possible to directly acquire the elements of an inverse system matrix, especially through the classification process. The feedforward architecture of a single-layer network is illustrated in Fig. 1. Spectrum vector $\mathbf{u}$ is used as the input data, and the corresponding voxel vector $\mathbf{c}$ becomes the target data. The input units are fully connected to the output layer nodes whose value is the weighting sum of the input values transformed by activation function $\sigma$ [18]. Considering no bias terms, the output value for the weighting matrix $\mathbf{W}$ is expressed as follows:

$$\mathbf{c} = \sigma(\mathbf{W}\mathbf{u}). \quad (6)$$

As compared to (4), the weighting matrix corresponds to the inverse system matrix $\mathbf{S^{-1}}$ when the linear activation function is used.

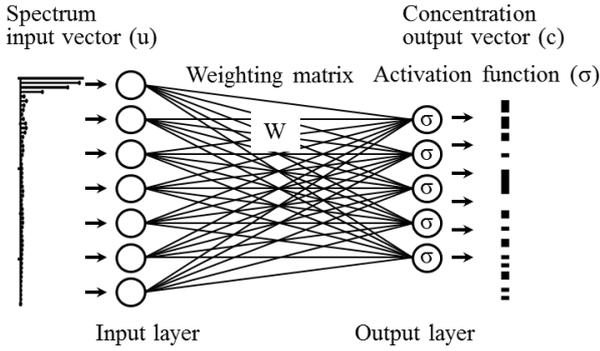

Fig. 1. The architecture of a single-layer neural network. The spectrum data and their corresponding voxel values are used as the training datasets.

Supervised learning is the process of finding the optimal values of the weight matrix elements. The network is trained using a generalized delta rule based on a gradient descent algorithm [19], where the weighting parameters are revised in a way that minimizes cost function $E$. The updating process of the weighting matrix at the respective iteration $k$ is given by

$$\mathbf{W}_{k+1} = \mathbf{W}_k - \eta \nabla E(\mathbf{W}_k), \tag{7}$$

where $\eta$ is the learning rate. After a successful training process, the networks can easily reconstruct the optimal image through the forward propagation of the spectral input vector.

A nonlinear activation such as a sigmoid function $\sigma_s(\mathbf{z})$ is generally used to execute the network learning:

$$\sigma_s(\mathbf{z}) = \frac{1}{1+\exp(-\mathbf{z})}. \tag{8}$$

The networks effectively classifies nonlinearly separable data into the target space. A kind of inverse kernel matrix can be obtained through the classification procedure, although it appears as an embedded form in a sigmoid function.

In the case of a multi-layer neural network with hidden layers, the inverse kernel is derived from the product of each layer's weight matrix. The classification property will be enhanced in this network type.

## III. Results of Image Reconstruction

### 1. Preparation of MPI Datasets

The MPI datasets were prepared by calculating the signal spectrum corresponding to the randomly distributed magnetic particles in a one-dimensional region. For the model-based simulation in (3), we adopt the spatial distribution of the superparamagnetic iron-oxide (SPIO) nanoparticle shown in Fig. 2(a), which is a representative material used as a contrast agent. The simulation parameters were set to allow for an actual operation of the MPI scanner [4]. A drive field with a frequency of 20 kHz and a gradient field of 2 T/m/$\mu_0$ was used to scan the trajectory. The scan range was determined from the ratio of the drive field amplitude to the gradient field intensity. Here, we set the drive field amplitude to 32 mT/$\mu_0$, and thus the simulated scanner system enables measurements of a particle distribution composed of 129 voxels within the range of ±16 mm. The resolution of the voxels becomes 0.25 mm.

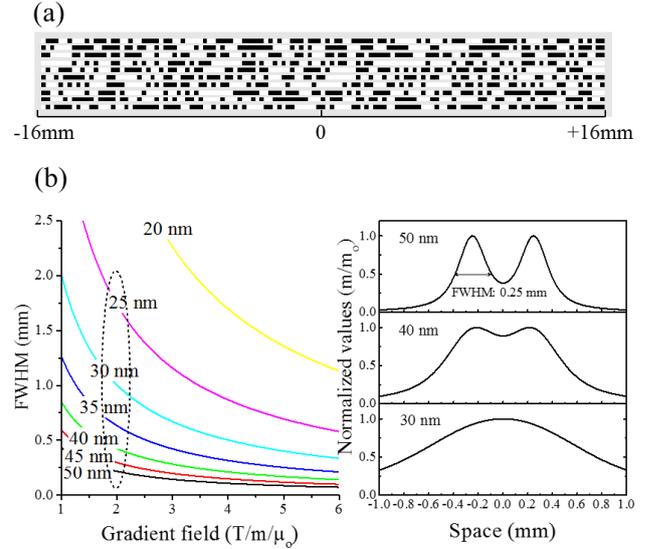

Fig. 2. Properties of convolved signals for (a) randomly distributed SPIO nanoparticles, and (b) the FWHM of the derivative of the Langevin function is plotted as a function of the nanoparticle size, where the image on the right shows a normalized convolved signal.

The size of the nanoparticles is a crucial factor in determining the spatial resolution of a reconstructed image. The Langevin function describes the magnetic response of the nanoparticles with respect to the applied magnetic field, as shown below:

$$M(H) = cmL\left(\coth(\alpha H) - \frac{1}{\alpha H}\right), \tag{9}$$

with $\alpha = \mu_0 m/k_B T$, where $k_B$ is a Boltzmann's constant and $T$ is the temperature. The magnetic moment $m$ of a spherical SPIO nanoparticle with a diameter $d$ and saturation magnetization $M_s = 0.6$ T/$\mu_0$ [4] is given by

$$m = M_s \pi d^3 / 6. \tag{10}$$

The full-width at half-maximum (FWHM) of the derivative of the Langevin function becomes a simple criterion for analyzing the image resolution. As shown in Fig. 2(b), the FWHM appears as an inversely proportional term to the diameter and gradient field, $G^{-1}d^{-1}$ [4]. The image on the right side of the figure shows the convolved signals for two adjacent particles. Two peaks can be sufficiently separated at the signal for a 50 nm particle, where the FWHM coincides with a voxel resolution of 0.25 mm.

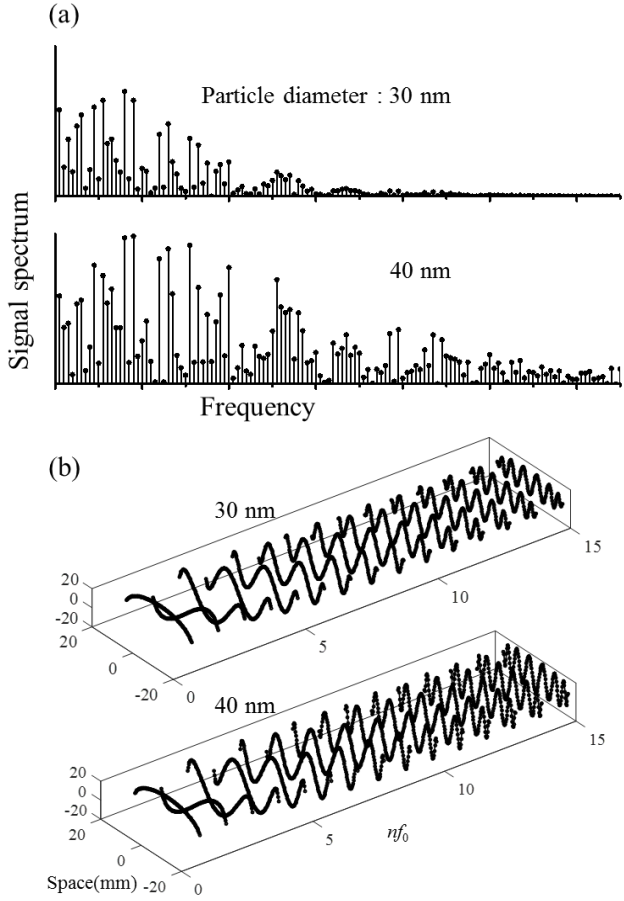

Fig. 3. (a) The corresponding spectrum data and (b) system functions of convolved signals for nanoparticles of 30 nm and 40 nm in sizes. The system function shows a larger convolution effect at a smaller particle size.

Figure 3 shows examples of signal spectrum data extracted for randomly generated nanoparticles with diameters of 30 nm and 40 nm. The particle concentration of each voxel has a binary value of zero or 1. The signal data in the frequency domain have higher harmonics of the base frequency, which results from the nonlinear characteristics of magnetization. The spectrum data in Fig. 3(a) show very different behaviors, where the data for 30 nm decrease rapidly with an increase in frequency, as compared to those for 40 nm. The system functions for these two data types are shown in Fig. 3(b). The component of the system function is interpreted as the convolution between the derivative of the magnetization curve and the Chebyshev polynomial component [6]. Larger convolution effects appear at smaller particle sizes. In general, the voxel information can be reconstructed using the system matrix and spectrum data through a regularization technique. This convolution property closely affects the spatial resolution of the retrieval image. We acquired the magnetic response signal of particles with various sizes to analyze the effect on the image retrieval through a neural network.

2. Single-layer Neural Network Image Reconstruction

Figure 4 shows the training behavior of single-layer neural networks with respect to the amount of training input data. The spectrum signals for 30,000 randomly distributed particle images with a nanoparticle size of 40 nm were prepared as the input training dataset. The number of input nodes was chosen to be 200 by considering a decrease in the spectrum values at the higher harmonic frequencies shown in Fig. 3(a). All spectrum data were normalized to effectively match the output value between zero and 1 of the sigmoidal activation function at each iteration step. Each initial weighting component has a random value, and the learning rate $\eta$ is set to 0.5 for convenience. The cost function uses the squared residuals. The mean squared error (MSE) is calculated from the subtraction of the forward output values to the known voxel values, which becomes a measure of the training performance for a neural network [18].

The MSE converges well with the increase in the number of iteration steps. As shown in Fig. 4(a), the minimum error is reached after a few hundred iteration steps for 30,000 training input data, which reaches close to $7.16 \times 10^{-5}$. We confirmed that more training data lead to a fast convergence during the learning process. Although the learning behavior with respect to many more iterations is not shown, the MSE gradually decreases by up to a low order of $10^{-5}$. We also observe that the high learning rate accelerates the convergence.

One thousand test datasets for a random particle distribution were prepared separately under the same conditions as the training data simulation. The test characteristics of the trained network are shown in Fig. 4(b). To clearly observe the level of quality, the reconstructed image displayed is from a portion of voxels for the ten initial samples. We found that it is possible to restore the voxel information of the original images more easily for a successfully trained neural network. The image pattern is apparently restored even for trained networks with 700 samples, whereas the MSE appears to be $5.69 \times 10^{-3}$, which reveals a partially insufficient retrieval of each pixel value. For 30,000 training datasets, MSE reaches $7.23 \times 10^{-5}$, and the difference from the original pixel image is indistinguishable by the naked eye. The quantity of the MSE for the testing is comparable to that of the training process.

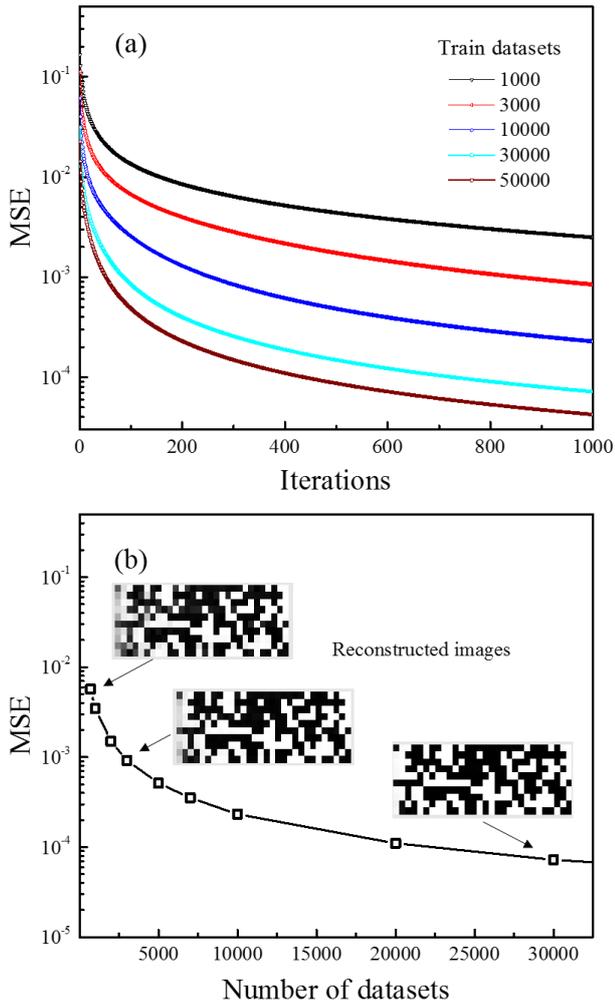

Fig. 4. The training and testing performances for a single-layer neural network: (a) convergence of MSE for the number of training datasets displayed during the learning process, and (b) each test MSE is comparable to that of the training process. The inset is shows a reconstructed image displaying a portion of voxels for the ten initial samples. The original image is clearly recovered for 30,000 training datasets.

The neural network performance depends strongly on the size of the superparamagnetic nanoparticle. Figure 5 shows the behavior of the learning process using 30,000 training datasets extracted from various sized nanoparticles. The training through a single-layer network is very poor for the datasets made of particles with a diameter of below 30 nm. This property should be closely associated with the convolution effect of the convolved signals shown in Fig. 2(b) because the spectrum is only a Fourier transform of a convolved signal. The target image used in our simulation was designed to have a pixel pitch of 0.25 mm. A 30 nm nanoparticle has an FWHM of approximately 1.0 mm, whereas it is difficult to discriminate two adjacent particles in a temporal signal. The resolving power significantly degrades with a decrease in particle size. We found that it is difficult to train a neural network using a spectral dataset with higher convolution effects.

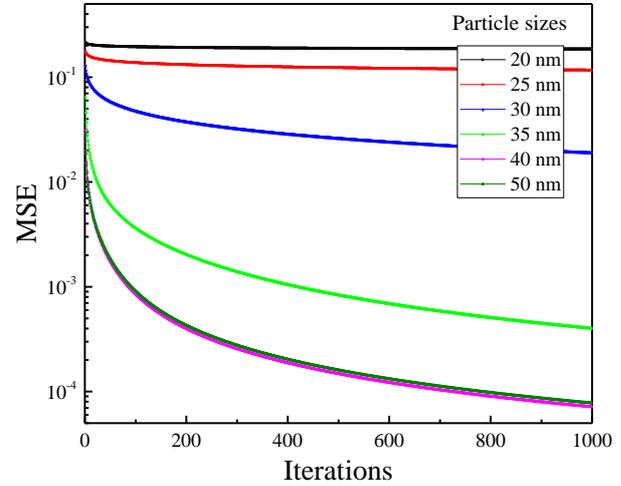

Fig. 5. The learning behavior of single-layer neural networks with respect to the sizes of the nanoparticles.

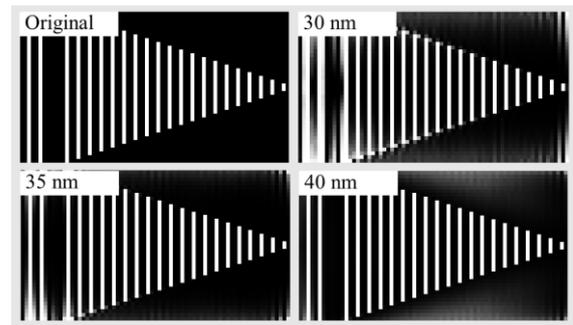

Fig. 6. Reconstructed images achieved through a trained single-layer neural network using the test datasets for a particular pattern image with various particle sizes. The 40 × 71 pixel images with about a half of repeated pattern in the direction of rows are displayed for convenience.

We test a particular pattern other than a randomized pixel image through trained neural networks. The test pattern image has a resolution of 40 × 129, where about one-half of a repeated pattern image is displayed. A total of 40 spectral datasets corresponding to each row pixel vector were prepared. The neural network technique can reconstruct an image easily through well-trained networks. A nearly original image is generated for trained networks with 40 nm sized particles. However, unlike a randomized pixel image, the network performance is slightly worse for image restoration for a pattern image with extremely low non-zero pixels.

## 3. Multi-layer Neural Network Image Reconstruction

We adopt feed-forward neural networks with one hidden layer. The convergence behavior is very different from that of a single-layer neural network, as shown in Fig 7(a). Each training curve has a particular threshold region, and then drops rapidly, where the hidden layer has 200 units. The MSE value for 30,000 training datasets falls drastically over several dozens of iterations to $10^{-6}$, and continues to decrease gradually up to $8.17 \times 10^{-7}$ for the final 1,000 steps. This value is two-orders of magnitude less than that for a single-layer network.

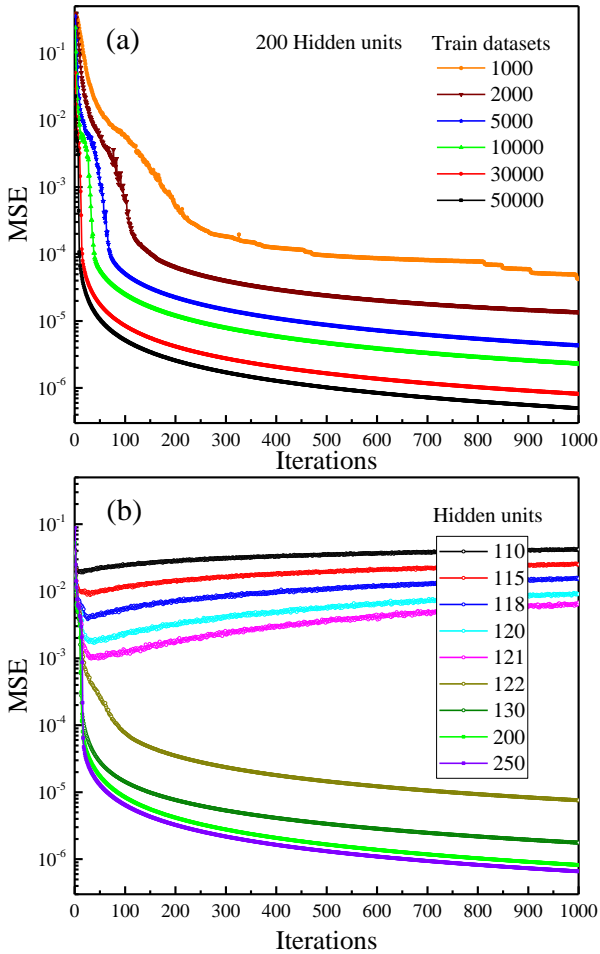

Fig. 7. The learning behavior of multi-layer neural networks with respect to the number of (a) training input datasets and (b) hidden units.

We also found that the learning capability is highly sensitive to the number of hidden units. A drastic change in training convergence appears for 121 hidden units, as shown in Fig. 7(b), and the learning performance is then gradually improved up to the saturation point. It should be noted that the network training undergoes a significant change within a similar range up to a 129 target-vector size. The training is difficult to achieve for a number of hidden units smaller than the length of the target vector. This might be related to acquiring a sufficient space of the hidden layer for a successful training.

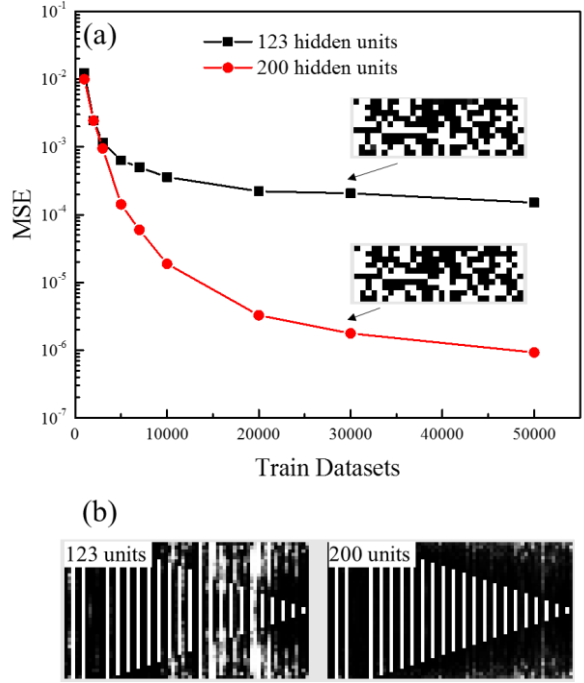

Fig. 8. The test properties for the trained multi-layer networks. (a) The MSE for each training dataset is the value after 1,000 iterations. (b) The testing of a particular pattern image is displayed.

The test performance for a single-layer neural network was comparable to that of its training process, as described in the previous section, whereas for the case of multi-layer neural networks, the test performance is relatively worse than the training capability. As shown in Fig. 8(a), the tested MSE value for the trained networks with 123 hidden units and 30,000 datasets appears to be about $2.06 \times 10^{-4}$, although the value for the training process reaches $6.19 \times 10^{-6}$, at which we are still able to retrieve the pixel pattern clearly. However, the networks with 200 hidden units reveal a test error of $1.77 \times 10^{-6}$, which is close to the value during the training procedure.

Figure 8(b) shows the test of a particular pattern through trained networks with 30,000 datasets. For 123 hidden units, the retrieved image is much worse than in a randomized pixel image, where we have trouble in recognizing an apparent pixel pattern. We observed that the test performance is improved as the number of hidden units increases, and thus, a network with 200 hidden units reconstructs a clear pattern image. Deterioration in quality still occurs for an image with an extremely low number of non-zero pixels.

Figure 9 shows the learning behavior for various sized nanoparticles. Differing from that in a single-layer network, great improvement is shown for 35 nm sized particles. This result is due to the enhanced classification function of the neural networks. That is, the weight matrices may imply the system environment for the MPI system. A complex architecture such as that of a deep neural network may lead to an improvement in network performance.

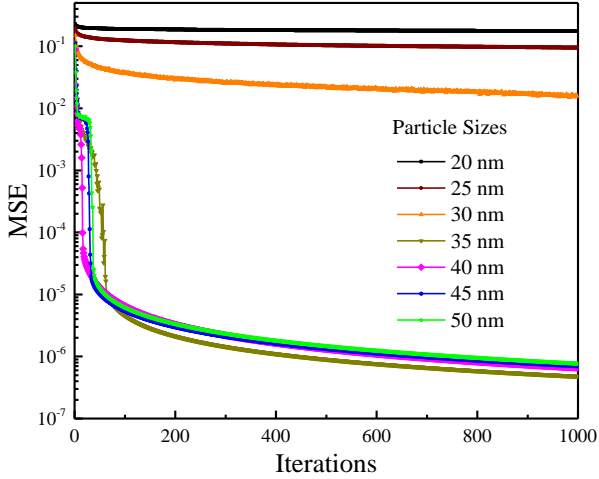

Fig. 9. The learning behavior of multi-layer neural networks with respect to the sizes of the nanoparticles. All the networks were learned by using 30,000 training datasets.

## IV. Discussions

1. Analysis of Inverse System Kernel

A successful training of a single-layer network means that the inverse kernel in the MPI can be found effectively through the classification process. Figure 10 shows the basis component of the $129 \times 200$ weighting matrix of our trained neural networks shown in Fig. 5. The first six column components of the weight matrix are displayed for convenience. The basis vectors appear in the form of Chebyshev polynomials.

As described in section II, weight matrix **W** is related to the inverse kernel. For the ideal particles in the MPI system, the system matrix is represented as the Chebyshev polynomials of the second kind multiplied by the weighting factor, $\sqrt{1-(Gx/A_D)^2}$ [6]. This function oscillates between constant extremums. In a real system, the magnitude of the polynomials at both boundary regions decreases smoothly because of the convolution property with the derivative of the magnetization curve, as shown in Fig. 3(b).

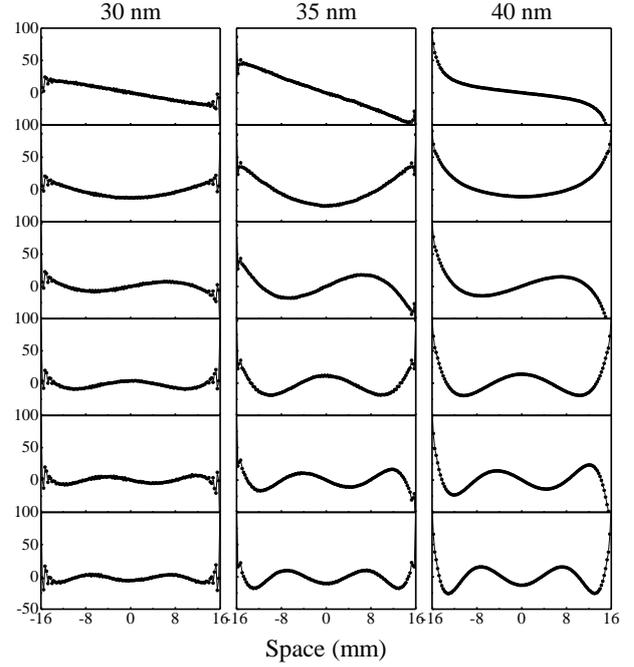

Fig. 10. Several column components of the $129 \times 200$ weight matrix of trained neural networks.

The Chebyshev polynomials $U_n$ satisfy the orthogonality condition [6]:

$$\int_{-1}^{1} U_n(z)U_m(z)\sqrt{1-z^2}\,dz = \frac{\pi}{2}\delta_{nm}. \quad (11)$$

Here, the $z$ variable indicates $Gx/A_D$. From the orthogonal relation in (11), the inverse matrix becomes the Chebyshev polynomials without a weighting factor. The weight matrix trained through datasets of 40 nm sized particles shows a clear shape of the basis components, as illustrated in Fig. 10. Each kernel function increases largely at both boundaries, which is a characteristic of the Chebyshev polynomials themselves. Furthermore, they reveal the orthogonal property. However, in the weighting matrix trained using smaller particles, the basis component does not illustrate a well-behaved curve, particularly at both ends. We found that the inverse kernel can be adequately obtained from well-trained neural networks.

Considering a linear activation function, the weighting matrix becomes an exact inverse system function. As illustrated in Fig. 11, the network shows a threshold for the training performance when a linear activation function is used. It has a limitation in acquiring an inverse matrix. On the other hand, under the sigmoidal activation function in previous results, a well-trained weight matrix consisting of a Chebyshev polynomial form can be searched more easily. It has been known that a nonlinearly separable activation function is very useful to classify the input data [18].

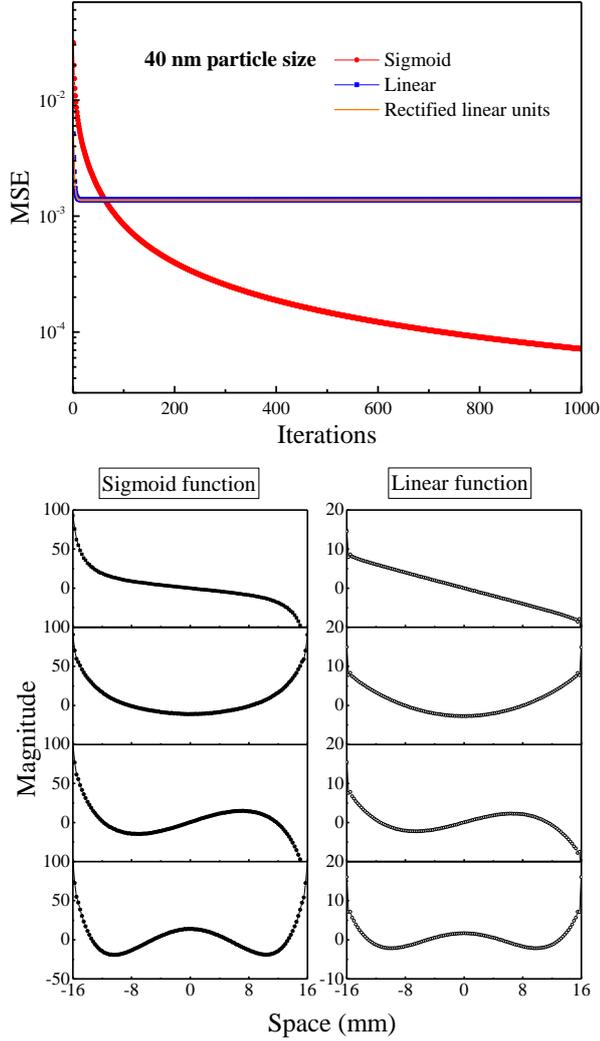

Fig. 11. The learning properties for single-layer neural networks with respect to the activation functions.

We interpret this phenomenon as a role of nonlinear target space. Equation (6) can be written using logit function:

$$\text{logit}(\mathbf{c}) = \mathbf{W}\mathbf{u} . \qquad (12)$$

The target space is nonlinear due to a logit function. The linear matrix equation for the MPI has a kernel with a large condition number, and it is therefore difficult to directly obtain an inverse matrix. However, in a nonlinear vector space of the output vector, the inverse matrix corresponding to the weighting matrix is obtainable. Strictly speaking, the weighting matrix $\mathbf{W}$ is not the exact inverse of system matrix $\mathbf{S}^{-1}$. As shown in Fig. 11, the curvature of the function is relatively different from that of the original polynomials. The values at both boundaries reveal a rapid increase, which is due to the nonlinear target space. However, this matrix acts as an inverse kernel in our system. A rigorous description will be applied later to analyze the effects of a nonlinear vector space on retrieving an inverse kernel.

2. Convolution Effects on Network Performance

The smaller nanoparticles result in the system function having largely convolved Chebyshev polynomials. When external noise is not considered, the resolution of the MPI is determined based on the deconvolution ability. A higher convolved polynomial leads to lower orthogonality, at which an ill-posed system is inevitable. An adequate regularization approach is important to solve this system [20]. In our system, the truncated SVD method retrieves the spatial information well for particles of above 30 nm.

This reconstruction property according to the particle size is similar to that of a neural network approach. Although a temporal signal of 30 nm particles seems unable to differentiate two adjacent particles, the datasets are trained through a nonlinear activation function. We found that the greater convolution effects impede the network training. This should be related to the incoherency of the system matrix [21], [22]. Its orthogonality worsens with a decrease in particle size. Namely, the training datasets extracted from a system function having higher incoherency are well trained through the single-layer neural networks. We also find that datasets with an external noise have a worse performance for the training process.

Multi-layer neural networks improves the property that classifies the input data into specific regions. In the MPI network system, the role of the weighting matrix may be beyond the simple system function that maps the spectral input data into pixel values. The weighting matrices are kernels having all of the network properties, where the classification is crucial. Our multi-layer neural networks show the better training performance. We found that, to enhance the performance of neural networks for the MPI system using large convolution effects, an optimization approach to the architecture of the networks is required. The neural networks, such as deep learning, accomplishes this purpose.

## V. Conclusions

Image reconstruction in the MPI system was successfully carried out using well-trained feed-forward neural networks. Networks having datasets with a relatively low convolution effect are well trained upon a nonlinear activation function. The learning process in single-layer neural networks is interpreted as a method for finding the inverse kernel elements under the incoherency conditions of the system matrix. A multi-layer neural network with one hidden layer shows potential for improving the training performance. This approach becomes a useful method for overcoming the computational reconstruction costs, and furthermore, an effective model for analyzing the internal structures such as the weighting matrix of neural networks.

## References

[1] B. Gleich and J. Weizenecker, "Tomographic Imaging using the Nonlinear Response of Magnetic Particles," *Nature*, vol. 435, no. 7046, Jun. 2005, pp. 1214–1217.

[2] J. Weizenecker, J. Borgert, and B. Gleich, "A Simulation Study on the Resolution and Sensitivity of Magnetic Particle Imaging," *Phys. Med. Biol.*, vol. 52, 2007, pp. 6363–6374.

[3] T. M. Buzug *et al.*, "Magnetic Particle Imaging: Introduction to Imaging and Hardware Realization," *Z. Med. Phys.*, vol. 22, no. 4, 2012, pp. 323−334.

[4] T. Knopp and T. M. Buzug, "Magnetic Particle Imaging," Springer-Verlag Berlin 2012.

[5] P. Goodwill et al., "Narrowband Magnetic Particle Imaging," *IEEE Trans. Med. Imag.*, vol. 28, no. 8, Aug. 2009, pp. 1231–1237.

[6] J. Rahmer et al., "Signal Encoding in Magnetic Particle Imaging: Properties of the System Function," *BMC Med. Imag.*, vol. 9, no. 1, 2009, pp. 4.

[7] T. Knopp et al., "Model-based Reconstruction for Magnetic Particle Imaging," *IEEE Trans. Med. Imag.*, vol. 29, no. 1, Jan. 2010, pp. 12–18.

[8] T. Sattel et al., "Single-sided Device for Magnetic Particle Imaging," *J. Phys. D: Appl. Phys.*, vol. 42, no. 1, 2009, pp. 1–5.

[9] K. Gräfe et al., "System Matrix Recording and Phantom Measurements with a Single Sided Magnetic Particle Imaging Device," *IEEE Trans. Magn.*, vol. 51, no. 2, 2015, pp. 65023031–3.

[10] K. Gräfe et al., "2D Images Recorded with a Single-sided Magnetic Particle Imaging Scanner," *IEEE Trans. Med. Imag.*, vol. 35, no. 4, Apr. 2016, pp. 1056–1065.

[11] T. Knopp *et al.*, "Weighted Iterative Reconstruction for Magnetic Particle Imaging," *Phys. Med. Biol.*, vol. 55, 2010, pp. 1577−1589.

[12] D. Boublil et al., "Spatially-adaptive Reconstruction in Computed Tomography using Neural Networks," *IEEE Trans. Med. Imag.*, vol. 34, no. 7, Jul. 2015, pp. 1474–1485.

[13] R. Cierniak, "A 2D Approach to Tomographic Image Reconstruction using a Hopfield-type Neural Network," *Artif. Intell. Med.*, vol. 43, 2008, pp. 113−125.

[14] C. E. Floyd, Jr., "An Artificial Neural Network for SPECT Image Reconstruction," *IEEE Trans. Med. Imag.*, vol. 10, no. 3, Sep. 1991, pp. 485−487.

[15] J. P. Kerr and E. B. Bartlett, "A Statistically Tailored Neural Network Approach to Tomographic Image Reconstruction," *Med. Phys.*, vol. 22, no. 5, 1995, pp. 601−610.

[16] D. C. Durairaj, M. C. Krishna, and R. Murugesan, "A Neural Network Approach for Image Reconstruction in Electron Magnetic Resonance Tomography," *Comput. Biol. Med.*, vol. 37, no. 10, Oct. 2007, pp. 1492−1501.

[17] T. Hatsuda et al., "Basic Study of Image Reconstruction Method Using Neural Networks with Additional Learning for Magnetic Particle Imaging," *International Journal on Magnetic Particle Imaging*, vol. 2, no. 2, Nov. 2016, pp. 1−5.

[18] C. M. Bishop, *Neural Networks for Pattern Recognition*, Oxford University Press, New York, 1995.

[19] D. E. Rumelhart, G. E. Hinton, and R. J. Williams, "Learning Representations by Back-propagating Errors," *Nature,* vol. 323, Oct. 1986, pp. 533−536.

[20] T. Knop and A. Weber, "Sparse Reconstruction of the Magnetic Particle Imaging System Matrix," *IEEE Trans. Med. Imag.*, vol. 32, no. 8, Aug. 2013, pp. 1473–1480.

[21] B. G. Chae and S. Lee, "Sparse-view CT Image Recovery using Two-step Iterative Shrinkage-thresholding Algorithm," *ETRI Journal*, vol. 37, no. 6, Dec. 2015, pp. 1251–1258.

[22] D. Donoho, "Compressed Sensing," *IEEE Trans. Inf. Theory*, vol. 52, no. 4, 2006, pp. 1289−1306.